\documentclass{sig-alternate-05-2015}
\input{macro}

\usepackage{multirow}
\usepackage{lscape}
\usepackage{subfigure}
\usepackage{url}
\usepackage{framed}
\usepackage{color}
\usepackage{balance}
\usepackage{colortbl}
\usepackage{nopageno}

\makeatletter
  \newcommand{\miniscule}{\@setfontsize\miniscule{3}{4}}
  \newcommand{\minisculee}{\@setfontsize\minisculee{4}{5}}
\makeatother

\begin{document}
\toappear{}
\setcopyright{acmcopyright}

\conferenceinfo{FSE '16}{Nov 13-19, 2016, Seattle, WA}

\acmPrice{\$15.00}

\title{A Large-Scale Empirical Comparison of Static and Dynamic Test Case Prioritization Techniques}
\author{
Qi Luo, Kevin Moran, and Denys Poshyvanyk\\
Department of Computer Science\\ College of William and Mary\\Williamsburg, Virginia 23185, USA\\
\{qluo,kpmoran,denys\}@cs.wm.edu}

\maketitle

\begin{abstract}
The large body of existing research in Test Case Prioritization (TCP) techniques, can be broadly classified into two categories: \textit{dynamic} techniques (that rely on run-time execution information) and \textit{static} techniques (that operate directly on source and test code).  Absent from this current body of work is a comprehensive study aimed at understanding and evaluating the static approaches and comparing them to dynamic approaches on a large set of projects.

In this work, we perform the first extensive study aimed at empirically evaluating four static TCP techniques comparing them with state-of-research dynamic TCP techniques at different test-case granularities (e.g., method and class-level) in terms of effectiveness, efficiency and similarity of faults detected. This study was performed on 30 real-word Java programs encompassing 431 KLoC. In terms of effectiveness, we find that the static call-graph-based technique outperforms the other static techniques at test-class level, but the topic-model-based technique performs better at test-method level. In terms of efficiency, the static call-graph-based technique is also the most efficient when compared to other static techniques. When examining the similarity of faults detected for the four static techniques compared to the four dynamic ones, we find that on average, the faults uncovered by these two groups of techniques are quite dissimilar, with the top 10\% of test cases agreeing on only $\approx$ 25\% - 30\% of detected faults.  This prompts further research into the severity/importance of faults uncovered by these techniques, and into the potential for combining static and dynamic information for more effective approaches.

\end{abstract}
\vspace{-0.1cm}


\section{introduction}
\label{sec:intro}

Modern software is constantly evolving; developers make various program changes to add new features or refactor existing code. During this process, it is crucial to ensure that developers do not introduce new bugs, known as software \textit{regressions}. {\em Regression testing} is a methodology for efficiently and effectively validating software changes against an existing test suite aimed at detecting such bugs \cite{Lu:ICSE16,Zhang:ICSM11}. One of the key tasks, which is of critical importance to the regression testing process, is regression test case prioritization.  

Regression test prioritization techniques reorder test executions in order to maximize a certain objective function, such as exposing faults earlier or reducing the execution time cost \cite{Lu:ICSE16}. For insatnce, Microsoft researchers have built test prioritization systems for development and maintenance of Windows for a decade~\cite{Srivastava:ISSTA02, Czerwonka:ICST11}. Moreover, a large body of research work has been proposed to design and evaluate regression TCP techniques~\cite{Walcott:06,zhang2013bridging,Rothermel:01,Rothermel:99, Lu:ICSE16, Korel:05}. Most traditional TCP techniques are based on the dynamic coverage information of the regression test suite in previous software versions. A typical dynamic approach applies a certain test prioritization strategy on a particular test coverage criterion to iteratively compute each test's priorty, and then ranks them to generate a prioritized list of tests. Researchers have proposed various approaches for traditional TCP, such as greedy (total and additional strategies)
~\cite{zhang2013bridging,Rothermel:01,Rothermel:99}, adaptive random testing ~\cite{Jiang:09}, and search-based strategies~\cite{Li:07}.

Although dynamic test prioritization techniques can be powerful in practice, they may not be always applicable due to notable disadvantages (e.g., time-consuming \cite{Mei:TSE12}, performance degradation over new program versions and test cases \cite{Lu:ICSE16}). Thus, researchers have proposed a number of test prioritization techniques that rely solely upon \textit{static} information extracted from the source and test code.  The introduction of purely static techniques begs several important questions in the context of past work on dynamic techniques, such as: \textit{How does the effectiveness of static and dynamic techniques compare on real-world software projects?} \textit{Do static and dynamic techniques uncover similar faults?}; \textit{How efficient are static techniques when compared to one another?}  The answers to these questions will guide future work in developing new test-case prioritization techniques.

Several empirical studies have been conducted in an attempt to examine and understand varying aspects of different white and black-box TCP approaches~\cite{Rothermel:99,Elbaum:TSE02,Do:04,Qu:ISSTA08,Thomas:EMSE14}.
However, there is a clear gap in the existing body of empirical studies characterized by the following: 1) recently proposed TCP techniques, particularly static techniques, have not been thoroughly evaluated against each other or against techniques that require dynamic coverage; 2) no previous study including static approaches comprehensively examines the impact of different test granularities (e.g., prioritizing entire test classes or individual test methods), the efficiency of the techniques, and the similarities in terms of uncovered faults; and 3) prior studies have not typically been conducted on several mature real-world software systems. Each of these points are important considerations that call for a thorough empirical investigation. Studying the effectiveness and similarity of faults uncovered for \textit{both} static and dynamic techniques would help inform researchers of potential opportunities to design more effective and robust TCP approaches.  Additionally, evaluation on a large group of sizable real-world java programs would help bolster the generalizability of these results.

To answer the previously posed questions and address the current gap in the existing body of TCP research we perform an extensive empirical study comparing four popular static TCPs, i.e., call-graph-based (with total and additional strategies) \cite{Zhang:ICSM09}, string-distance-based \cite{Ledru:ASE12}, and topic-model based techniques \cite{Thomas:EMSE14} to four state-of-the-art dynamic TCPs (i.e., the greedy-total \cite{Rothermel:99}, greedy-additional \cite{Rothermel:99}, adaptive random \cite{Jiang:09}, and search-based techniques \cite{Li:07}) on 30 real-world software systems. All TCPs are implemented based on the papers that initially proposed them and the implementation details are given in Section \ref{sec:impl}.  It is worth noting that different granularities of dynamic coverage may impact the effectiveness of dynamic TCPs. In this paper, we chose to examine statement-level coverage for dynamic techniques, since previous work \cite{Lu:ICSE16,Mei:TSE12} has shown that statement-level coverage is \textit{at least as effective} as other common coverage criteria (e.g., method and branch coverage) in the TCP domain. In our evaluation criteria we examine the effectiveness of these techniques, in terms of Average Percentage of Faults Detected (APFD), and the similarity of detected of faults at different test granularities (e.g., both method and class levels). Additionally we examine the efficiency, in terms of execution time, of static TCPs to better understand the time cost associated with running these approaches on subjects.

When examining static approaches, we found that the call-graph-based (with ``additional" strategy) technique outperforms all studied techniques at the test-class level. At the test-method level, the topic-model based technique performs better than other static techniques, but worse than two dynamic techniques, the additional and search-based techniques.  \textit{Our results indicate that the test granularity dramatically impacts the effectiveness of TCP techniques}. While nearly all techniques perform better at method-level granularity, the static techniques perform comparatively worse to dynamic techniques at method-level as opposed to class level. In terms of execution time, call-graph based techniques are the most efficient of the static TCPs. Finally, the results of our similarity analysis study suggest that there is minimal overlap between the uncovered faults of the studied dynamic and static TCPs, with the top 10\% of prioritized test-cases only sharing $\approx$ 25\% - 30\% of uncovered faults. This suggests that future TCPs may benefit from the severity/importance of faults uncovered by different techniques, and the potential for combining static and dynamic information.
This paper makes the following contributions:
\vspace{-0.18cm}
\begin{itemize}
\item To the best of the author's knowledge, this is the first extensive empirical study that compares the effectiveness, efficiency, and similarity of uncovered faults of both static \textit{and} dynamic TCP techniques at different granularities on a large set of real-world programs.
\vspace{-0.2cm}
\item{We discuss the relevance and potential impact of the findings in the study, and provide a set of learned lessons to help guide future research.}
\vspace{-0.2cm}
\item We provide a publicly available, extensive online appendix and dataset of the results of this study to ensure reproducibility and aid future research~\cite{Qi:FSE16}.
\end{itemize}
\vspace{-0.2cm}

\vspace{-0.25cm}
\section{Background \& Related Work}
\vspace{-0.1cm}
\label{sec:techniques}
In this section we formally define the TCP problem, introduce the studied techniques in the context of the related work, and distill the novelty and research gap that our proposed study fulfills.
Rothermel et al.~\cite{Rothermel:01} formally defined the
test prioritization problem as finding $T'\in P(T)$, such that $\forall
T'', T''\in P(T)\wedge T''\neq T'\Rightarrow f(T')\ge f(T'')$, where $P(T)$
denotes the set of permutations of a given test suite $T$, and $f$
denotes a function from $P(T)$ to real numbers. In this section, we introduce the underlying methodology of our studied techniques in detail, which include static TCP techniques (Section~\ref{sec:static-pri}) and dynamic TCP techniques (Section~\ref{sec:dynamic-pri}). All studied techniques attempt to address the TCP problem formally enumerated above with the objective function of uncovering the highest number of faults with the fewest number of test cases. In our study, we limit our focus to static TCPs that require only source code and test cases, and the dynamic ones that only require dynamic coverage and test cases as inputs for two reasons: 1) this represents fair comparison of similar techniques that leverage traditional inputs (e.g., test cases, source code and coverage info), and 2) the inputs needed by other techniques (e.g., requirements, code changes, user knowledge) are not always available in real-world subject programs. Additionally, we discuss existing empirical studies (Section~\ref{sec:study-pri}).  We discuss our own re-implementation of these tools later in Section 3.
\vspace{-0.42cm}
\subsection{Static TCP Techniques}
\vspace{-0.03cm}
\label{sec:static-pri}

\noindent{\bf Call-Graph-Based.}
This technique builds a call graph for each test case to obtain a set of transitively invoked methods, called \textit{relevant methods}~\cite{Zhang:ICSM09}.  The test cases with a higher number of relevant methods in the call-graph are treated as the ones with higher test ability and thus are prioritized first. This approach encompasses two sub-strategies, the \textit{total} strategy prioritizing the test cases with higher test abilities earlier, and the \textit{additional} strategy prioritizing the test cases with higher test abilities excluding the methods that have already been covered by the prioritized test cases. Mei \emph{et al.} extended this work to measure the test abilities of the test cases using the number of statements in their relevant methods instead of the number of relevant methods~\cite{Mei:TSE12}. The intuition here is that test cases with a larger number of statements in their relevant methods are more likely to detect faults. As defined in previous work \cite{Henard:ICSE16, Thomas:EMSE14}, a white-box static approach requires access to \textit{both} the source code of subject programs, and other types of information (e.g., test code), whereas black-box static techniques do not require the source code of subject programs. Static techniques can be classified as either white or black box \cite{Thomas:EMSE14}, whereas most dynamic techniques (including the ones considered in this study) are considered white-box techniques since they require access to the subject system's source code. Thus the call-graph based technique is classified as a white box approach, whereas the other two studied techniques are black-box approaches. We consider both types of static techniques in this paper in order to thoroughly compare them to a set of techniques that require dynamic computation of coverage.

\noindent{\bf String-Distance-Based.} The key idea underlying this technique is that test cases that are most different from already executed test cases, as measured by textual similarity based on string-edit distance, should be prioritized earlier \cite{Ledru:ASE12}. The reason is that test cases that are textually dissimilar have a higher probability of executing different code paths and thus detecting more bugs. This technique is a \textit{black-box} static technique since it uses only the test case code. In this technique, four string-edit distances are introduced to calculate the gap between each pair of test cases: Hamming, Levenshtein, Cartesian, and Manhattan distances. Based on prior experimental results~\cite{Ledru:ASE12}, Manhattan distance performs best in terms of detecting faults. Thus, in our study, we implemented the string-based TCP based on the paper by Ledru \emph{et al.} \cite{Ledru:ASE12}, and chose Manhattan distance as the representative string distance computation for this technique. We provide explicit details regarding our implementation of studied techniques in Section \ref{sec:study}.

\noindent{\bf Topic-Based.} This static black-box technique uses semantic-level topic models to represent tests of differing functionality, and gives higher prioritization to test cases that contain different topics form those already executed \cite{Thomas:EMSE14}. The authors claim that topics, which abstract test cases' functionality, can capture more information than the existing static test prioritization techniques, and is robust in obtaining differences between test cases. \Comment{For each test case, its corresponding code is treated as a basic document.}The technique creates a vector for the code of each test case, including the test case's correlation values with each topic. After that, it calculates the distances between test cases using Manhattan distance, and defines the distance between one test case and a set of test cases as the minimum distance between this test case and all test cases in the set. During prioritization process, the test case which is farthest from all other test cases is selected and put into the prioritized set. Then, the technique iteratively selects the test cases that are farthest from the prioritized set. We implemented this technique based on the original paper \cite{Thomas:EMSE14} with the same described parameters.

\noindent{\bf Other Approaches.} An approach presented by Jiang \emph{et al.} calculates the distances between test cases based on the input values of test cases, and favors the test cases which are farthest from the already prioritized test case set~\cite{Jiang:compsac13}. Recently, Saha \emph{et al.} proposed an approach that uses Information Retrieval (IR) techniques to recover the traceability links between test cases and code changes, and sorts the test cases based on their relevant code changes, with those having more relevant code changes being prioritized first~\cite{Saha:ICSE15}. These techniques require additional information, such as the test input and code changes. Recall that, we focus on automated TCPs that require only the source code and the test cases of the subjects. Thus, we choose the call-graph-based, string-based and topic-based techniques as the focus.
\vspace{-0.12cm}
\subsection{Dynamic TCP Techniques}
\label{sec:dynamic-pri}
\vspace{-0.05cm}

\noindent{\bf Greedy Techniques.}
As explained in our overview of the Call-Graph-based approach, traditional dynamic TCPs use two sub-techniques, the \textit{total} strategy and \textit{additional} strategy, to prioritize test cases based on the code coverage. Similarly, the total strategy prioritizes test cases based on their code coverage, and the additional strategy prioritizes test cases based on their code coverage excluding the code elements that have been covered by prioritized test cases. Thus, the total strategy favors the test cases that cover more code, but the additional strategy would select the test cases that can cover different code from the already prioritized test cases earlier. In our study, we implemented the greedy techniques based on the work by Rothermel et al. \cite{Rothermel:99}. The \textit{additional} strategy of this approach has been widely considered as one of most effective TCPs in previous works~\cite{Jiang:09,zhang2013bridging}. Recently, Zhang \emph{et al.} proposed a novel approach to bridge the gaps between these two strategies by unifying the strategies based on the fault detection probability~\cite{zhang2013bridging,Hao:TOSEM14}.

Different code coverage criteria, such as statement coverage \cite{Rothermel:99}, basic block and method coverage~\cite{Do:04}, Fault-Exposing-Potential (FEP) coverage~\cite{Elbaum:TSE02}, transition and round-trip coverage~\cite{Xu:ICST10}, have been investigated in the domain of dynamic TCP.  For instance, Do \emph{et al.} use both method and basic block coverage information to prioritize test cases~\cite{Do:04}. Elbaum \emph{et al.} proposed an approach that prioritizes test cases based on their FEP and fault index coverage~\cite{Elbaum:TSE02}, in which the test cases exposing more potential faults will be assigned a higher priority.
Kapfhammer \emph{et al.} use software requirement coverage to measure the test abilities of test cases for test prioritization ~\cite{Kapfhammer:2007}.

\noindent{\bf Adaptive Random Testing.}
Jiang \emph{et al.} proposed a novel approach, called Adaptive Random Test Case Prioritization (ART), which introduces the adaptive random testing strategy~\cite{Chen:04} into the TCP domain~\cite{Jiang:09}. ART first randomly selects a set of test cases iteratively to build a candidate set, then it selects from the candidate set the test case farthest away from the prioritized set. The whole process is repeated until all test cases have been selected. To find the farthest test case, ART first calculates the distance between each pair of test cases using Jaccard distance based on their coverage, and then calculates the distance between each candidate test case and the prioritized set. Three types of distances are used to calculate the distance between one test case and the prioritized set, {\em min}, {\em avg} and {\em max}. For example, {\em min} is the minimum distance between the test case and the prioritized test case. The authors compared ART with different distances and also the random TCP technique. The results show that ART with {\em min} distance performs best. Thus, in our empirical study, we implemented ART based on Jiang \emph{et al.}'s paper \cite{Jiang:09} and chose {\em min} distance to estimate the distance between one test case and the prioritized set.

\noindent{\bf Search-based Techniques.}
Search-based TCP techniques introduce the meta-heuristic search algorithm into the TCP domain, exploring space of test case combinations to find the ranked list of test cases that detect faults more quickly~\cite{Li:07}. Li \textit{et al.} propose two search-based test prioritization techniques, hill-climbing-based and genetic-based. The hill-climbing-based technique evaluates all neighbors, locally searching the ones that can achieve largest increase in fitness. The genetic technique halts evolution when a predefined termination condition is met, e.g., the fitness function value reaches a given value or a maximal number of iterations has been reached. Our empirical study uses the genetic-based test prioritization approach as the representative search-based test case prioritization technique, because previous results demonstrate that genetic-based technique is more effective in detecting faults~\cite{Li:07}.

\noindent{\bf Other Approaches.} Several other techniques based on leveraging dynamic program information have been proposed, but do not fit into any of the classifications enumerated above \cite{Islam:CSMR12,Tonella:ICSM06,Nguyen:ICWS11}. Islam \emph{et al.} presented an approach that recovers traceability links between system requirements and test cases using (IR) techniques, and dynamic information such as execution cost and code coverage, to prioritize test cases~\cite{Islam:CSMR12}. \Comment{Tonnella \emph{et al.} proposed a machine learning algorithm, called Case-Based Ranking, to prioritize test cases by combining several pieces of information including user knowledge, code coverage, and fault proneness~\cite{Tonella:ICSM06}. }Nguyen \emph{et al.} designed an approach that uses IR techniques to recover the traceability links between change descriptions and execution traces for test cases to identify the most relevant test cases for each change description~\cite{Nguyen:ICWS11}. However, these TCPs require more information (e.g., execution cost, user knowledge, code changes) than code coverage. In this paper, we choose dynamic techniques that require only code coverage and test cases for comparison, thus we select three techniques (i.e., Greedy (with total, additional strategies), ART, and Search-based). Recall that we do not study the potential impact of coverage granularity on the effectiveness of dynamic TCPs. Previous work has already shown that statement-level coverage is \textit{at least as effective} as other coverage types \cite{Lu:ICSE16,Mei:TSE12}, thus we chose statement-level coverage for all studied dynamic TCPs.
\vspace{-0.15cm}
\subsection{Empirical studies on TCP techniques}
\vspace{-0.05cm}
\label{sec:study-pri}
Several studies empirically evaluating TCP techniques~\cite{Kasurinen:2010, Rothermel:99, Catal:2013, Wang:2014, do06nov, Epitropakis:2015,You:2011,Smith:SAC09, Henard:ICSE16,Lu:ICSE16,Elbaum:SQI04,Elbaum:TSE02,Yoo:STVR12,Epitropakis:2015,Qu:ISSTA08} have been published.  In this subsection we discuss the studies most closely related to our own in order to illustrate the novelty and research gap filled by our proposed study. Rothermel \emph{et al.} conducted a study for unordered, random, and dynamic TCP techniques (e.g., coverage based, FEP-based) on C programs, to evaluate their abilities of fault detection~\cite{Rothermel:99}. Elbaum \emph{et al.} conducted a study for several dynamic TCP techniques on C programs, to evaluate the impact of program versions, program types and different code granularity on the effectiveness of TCP techniques~\cite{Elbaum:TSE02}. \Comment{Additionally, they conducted another study for coverage-based techniques on eight C programs to evaluate the impact of the program attributes, test suites and modifications on the effectiveness of fault detection~\cite{Elbaum:SQI04}. }Thomas et. al \cite{Thomas:EMSE14} compared the topic-based technique with the string-based and call-graph-based techniques and the greedy-additional dynamic technique at method-level on two subjects. However, this study is limited by a small set of subject programs, a comparison to only one dynamic technique, a comparison only at method-level, and no investigation of fault detection similarity among the approaches.

\Comment{Qu~\cite{Qu:ISSTA08} conducted an empirical study on seven versions of Vim to analyze the impact of software configurations on the effectiveness of several dynamic test prioritization techniques.} Do \emph{et al.} presented a study of dynamic test prioritization techniques (e.g., random, optimal, coverage-based) on four Java programs with JUnit to demonstrate that these techniques can be effective not only on C but also on Java programs, but different languages and testing paradigms may lead to divergent behaviors~\cite{Do:04}. They also proposed an empirical study to analyze the effects of time constraints on TCP techniques~\cite{Do:08}. Henard \emph{et al.} recently conducted a study comparing white and black-box TCP techniques in which the effectiveness, similarity, efficiency, and performance degradation of several techniques was evaluated.  While this is one of the most complete studies in terms of evaluation depth, it does not consider the static techniques considered in this paper. Thus, our study is differentiated by the unique goal of understanding the relationships between purely static and dynamic TCPs.

To summarize, while each of these studies offers valuable insights, none of them provides an in-depth evaluation and analysis of the effectiveness, efficiency, and similarity of detected faults for static TCP techniques and comparison to dynamic TCP techniques on a set of mature open source software systems.  This illustrates that a clear research gap exists in prior work empirically comparing more traditional techniques based on dynamic information against those that operate purely on static code artifacts.  The study conducted in this paper is meant to close this gap, and offer researchers and practitioners insight into the similarity and trade-offs between such approaches.

\vspace{-0.2cm}
\section{Empirical Study}
\label{sec:study}
In this section, we state our research questions, and enumerate the subject programs, test suites, study design, and implementation of studied techniques in detail.
\subsection{Research Questions (RQs):}
\vspace{-0.2cm}

\begin{itemize}
\item [\textbf{RQ$_1$}] How do static TCP techniques compare with each other and with dynamic techniques in terms of \textit{effectivness} measured by APFD?     \vspace{-0.2cm}
\item [\textbf{RQ$_2$}] How does the test granularity impact the effectiveness of both the static and dynamic TCP techniques?
    \vspace{-0.2cm}
\item [\textbf{RQ$_3$}] 
How \textit{similar} are different TCP techniques in terms of detected faults?
\vspace{-0.2cm}
\item [\textbf{RQ$_4$}] How does the \textit{efficiency} of static techniques compare with one another in terms of execution time cost? 
    \vspace{-0.2cm}
\end{itemize}
To aid in answering \textbf{RQ$_1$}, we introduce the following null and alternative hypotheses. The hypotheses are evaluated at a 0.05 level of significance:
\vspace{-0.2cm}
\begin{description}
\item[\textbf{$H_{0}$}:] There is no statistically significant difference in the effectiveness between the studied TCPs.
\vspace{-0.2cm}
\item[\textbf{$H_{1}$}:] There is a statistically significant difference in the effectiveness between the studied TCPs.
\vspace{-0.2cm}
\end{description}\begin{table}
\tiny
\vspace{-0.2cm}
\setlength{\tabcolsep}{4.5pt}
\center\caption{\label{tab:sub}\small The stats of the subject programs: Size: \#Loc; TM: \#test cases at method level; TC: \#test cases at class level; All: \#all mutation faults; Detected: \#faults can be detected by test cases.}\vspace{-0.1cm}
\begin{tabular}{|l||c|c|c|c|c|}
\hline
Subject Programs&Size&\#TM&\#TC&{Detected}&All\\
\hline\hline
P1-Java-apns&3,234&87&15&412&1,122\\
P2-gson-fire&3,421&55&14&847&1,064\\
P3-jackson-datatype-guava&3,994&91&15&313&1,832\\
P4-jackson-uuid-generator&4,158&45&6&802&2,039\\
P5-jumblr&4,623&103&15&610&1,192\\
P6-metrics-core&5,027&144&28&1,656&5,265\\
P7-low-gc-membuffers&5,198&51&18&1,861&3,654\\
P8-xembly&5,319&58&16&1,190&2,546\\
P9-scribe-java&5,355&99&18&563&1,622\\
P10-gdx-artemis&6,043&31&20&968&1,687\\
P11-protoparser&6,074&171&14&3,346&4,640\\
P12-webbit&7,363&131&25&1,268&3,833\\
P13-RestFixture&7,421&268&30&2,234&3,278\\
P14-LastCalc&7,707&34&13&2,814&6,635\\
P15-lambdaj&8,510&252&35&3,382&4,341\\
P16-javapoet&9,007&246&16&3,400&4,601\\
P17-Liqp&9,139&235&58&7,962&18,608\\
P18-cassandra-reaper&9,896&40&12&1,186&5,105\\
P19-raml-java-parser&11,126&190&36&4,678&6,431\\
P20-redline-smalltalk&11,228&37&9&1,834&10,763\\
P21-jsoup-learning&13,505&380&25&7,761&13,230\\
P22-wsc&13,652&16&8&1,687&17,942\\
P23-rome&13,874&443&45&4,920&10,744\\
P24-JActor&14,171&54&43&132&1,375\\
P25-jprotobuf&21,161&48&18&1,539&10,338\\
P26-worldguard&24,457&148&12&1,127&25,940\\
P27-commons-io&27,263&1125&92&7,630&10,365\\
P28-asterisk-java&39,542&220&39&3,299&17,664\\
P29-ews-java-api&46,863&130&28&2,419&31,569\\
P30-joda-time&82,998&4,026&122&20,957&28,382\\
\hline\hline
Total&431,329&8,959&845&92,797&257,807\\\hline
\end{tabular}
\vspace{-0.5cm}
\end{table} 

\subsection{Subject Programs, Test Suites and Faults}
We conduct our study on 30 real-world Java programs. The program names and sizes in terms of lines of code (LOC) are shown in Table~\ref{tab:sub}, where the sizes of subjects vary from 3,234 to 82,998 LoC; all are available on GitHub\cite{github}. Our subjects are larger in size and quantity than previous work in the TCP domain~\cite{Lu:ICSE16,Henard:ICSE16,Thomas:EMSE14,Ledru:ASE12,Jiang:09}. To perform this study, we checked out the most current master branch of each program, and provide the version IDs in our online appendix \cite{Qi:FSE16}. For each program, we used the original JUnit test suites for the corresponding program version. Since one of the goals of this study is to understand the impact of test granularity on the effectiveness of TCP techniques, we introduce two groups of experiments in our empirical study based on two test-case granularities: (i) the test-method and (ii) the test-class granularity. The numbers of test cases on test-method level and test-class level are shown in Columns 3 \& 4 of Table~\ref{tab:sub} respectively.

One goal of this empirical study is to compare the effectiveness of different test prioritization techniques by evaluating their fault detection capabilities. Thus, each technique will be evaluated on a set of program faults introduced using mutation analysis. As mutation analysis has been widely used in regression test prioritization evaluations~\cite{zhang2013bridging,Do:05,Lu:ICSE16,Zhang:ISSTA13} and has been shown to be suitable in simulating real program faults~\cite{Andrews:05,Just:FSE14}, this is a sensible method of introducing program defects.  We applied the PIT~\cite{PIT} mutation tool's built-in mutators to produce mutation faults for each project.  All mutation operators can be found in our online appendix~\cite{Qi:FSE16}. Note that not all produced mutation faults can be detected/covered by test cases, thus we ran PIT with all test cases to obtain the faults that can be detected and used these detected faults in our study. The numbers of detected mutation faults and the numbers of all mutation faults are shown in Columns 5 and 6 of Table~\ref{tab:sub} respectively. As the table shows, the numbers of detected mutation faults range from 132 to 20,957. There are of course certain threats to validity introduced by such an analysis, namely the the potential bias introduced by the presence of equivalent and trivial mutants \cite{Andrews:TSE06,Ammann:ICST14}. We summarize the steps we take in our methodology to mitigate this threat in Section \ref{sec:threats}.
\vspace{-0.15cm}
\subsection{Design of the Empirical Study}
\vspace{-0.05cm}
As discussed previously (see Section \ref{sec:techniques}), we limit the focus of this study to TCP techniques that do not require non-traditional inputs, such as code changes or software requirements. We select two white-box and two black-box static techniques, and four white-box dynamic techniques with statement-level coverage as the subject techniques for this study, which are listed in Table~\ref{tab:techniques}. We sample from both white and black box approaches as the major goal of this study is to examine the effectiveness and trade-offs of static and dynamic TCPs under the assumption that both the source code of the subject application, as well as the test cases are available. It is worth noting that our evaluation employs \textit{two} versions of the static topic model-based technique, as when contacting the authors of \cite{Thomas:EMSE14}, they suggested that an implementation using the Mallet \cite{Mallet} tool would yield better results than their initial implementation in R \cite{Thomas:EMSE14}. \Comment{There are various coverage granularities for the dynamic techniques, such as statement-level, method-level and class-level. Previous research showed that statement-level TCP techniques perform best~\cite{Mei:TSE12,Hao:TOSEM14}, thus, in our study, we choose statement-level coverage for the dynamic TCP techniques.}We now describe the experimental procedure utilized to answer each RQ posed above.

\begin{table}
\vspace{-0.2cm}
\center
\setlength{\tabcolsep}{4.6pt}
\scriptsize
\caption{\label{tab:techniques}Studied TCP Techniques}
\hspace{-0.15cm}
\begin{tabular}{|l||c|c|}
\hline
Type&Tag&Description\\
\hline
\hline
\multirow{4}{*}
{Static}&$TP_{cg-tot}$&Call-graph-based (total strategy)\\
&$TP_{cg-add}$&Call-graph-based  (additional strategy)\\
&$TP_{str}$&The string-distance-based\\
&$TP_{topic-r}$&Topic-model-based using R-lda package\\
&$TP_{topic-m}$&Topic-model-based using Mallet\\
\hline
\hline
\multirow{4}{*}
{Dynamic}&$TP_{total}$&Greedy total  (statement-level)\\
&$TP_{add}$&Greedy additional (statement-level)\\
&$TP_{art}$&Adaptive random (statement-level)\\
&$TP_{search}$&Search-based (statement-level)\\
\hline
\end{tabular}
\vspace{-0.6cm}
\end{table}

\textbf{RQ$_1$:} The \textit{goal} of \textbf{RQ$_1$} is to compare the effectiveness of different TCP techniques, by evaluating their fault detection capabilities. Following existing work~\cite{zhang2013bridging,Lu:ICSE16}, we fixed
the number of faults for each subject program. That is, we randomly chose 500 different mutation faults and partitioned the set of all faults into groups of five (e.g., a mutant group) to simulate each faulty program version. Thus, 100 different faulty versions (i.e., 500/5 = 100) were generated for each program. If a program has less than 500 mutation faults, we use all detected mutation faults for this program and separate these faults into different groups (five faults per group).  For the static techniques, we simply applied the techniques as described in Sections \ref{sec:techniques} \& \ref{sec:impl} to the test and source code of each program to obtain the list of prioritized test cases for each mutant group. For the dynamic techniques, we obtained the coverage information of the test-cases for each program.  We then used this coverage information to implement the dynamic approaches as described in Sections \ref{sec:techniques} \& \ref{sec:impl}.  Then we are able to collect the fault detection information for each program according to the fault locations.

To measure the effectiveness in terms of rate of fault detection for each studied test prioritization technique, we utilize the well-accepted Average Percentage of Faults Detected (APFD) metric in TCP domain~\cite{Rothermel:99, Zhang:ICSM09, Rothermel:01, Do:06, Elbaum:00, Elbaum:TSE02, Elbaum:2003}. Formally speaking, let $T$ be a test suite and $T'$ be a permutation of $T$, the APFD metric for $T'$ is computed according to the following metric:
\vspace{-0.1cm}
\begin{equation}\label{Equ:APFD}
    APFD=1- \frac{\sum_{i=1}^{m}TF_i}{n*m}+\frac{1}{2n}
    \vspace{-0.1cm}
\end{equation}
where $n$ is the number of test cases in $T$, $m$ is the number of faults, and $TF_i$ is the position of the first test case in $T'$ that detects fault $i$. Recall that every subject program has 100 mutant groups (five mutations per group). Thus, we created 100 faulty versions for each subject (each version contains five mutations) and ran all studied techniques over these 100 faulty versions. That is, running each technique 100 times for each subject. Then, we performed statistical analysis based on the APFD results of these 100 versions. To test whether there is a statistically significant difference between the effectiveness of different techniques, we first performed a one-way ANOVA analysis on the mean APFD values for all subjects and a Tukey HSD test\cite{Tukey}, following the evaluation procedures utilized in related work \cite{Mei:TSE12,Lu:ICSE16}. The ANOVA test illustrates whether there is a statistically significant variance between all studied techniques and the Tukey HSD test further distinguishes techniques that are significantly different from each other, as it classifies them into different groups based on their mean APFD values~\cite{Tukey}. These statistical tests give a statistically relevant overview of whether the mean APFD values for the subject programs differ significantly.  Additionally, we performed a Wilcoxon signed-rank test between each pair of TCP techniques for their average APFD value across all subject techniques, to further illustrate the relationship between individual subject programs. We choose to include this non-parametric test since we cannot make assumptions about wether or not the data under consideration is normally distributed.

\textbf{RQ$_2$:} The \textit{goal} of this \textbf{RQ} is to analyze the impact of different test granularities on the effectiveness of TCP techniques. Thus, we choose two granularities: test-method and test-class level. The test-method level treats each JUnit test method as a test case, while test-class level treats each JUnit test class as a test case.  We examine both the effectiveness and similarity of detected faults for both granularities.

\textbf{RQ$_3$:} The \textit{goal} of this \textbf{RQ} is to analyze the similarity of detected faults for different techniques to better understand the level of equivalency of differing strategies. It is clear that this type of analysis is important, as while popular metrics such as APFD measure the \textit{effectiveness} between two different techniques, this does not reveal the similarity of the test cases in terms of uncovered faults. For instance, let us consider two TCP techniques A and B.  If technique A achieves an APFD of $\approx 60\%$ and technique B achieves an APFD of $\approx 20\%$, while this gives a measure of relative effectiveness, the APFD does not reveal how similar or orthogonal the techniques are in terms of the faults detected.  For instance, all of the faults uncovered by top ten test cases from technique B could be different than those discovered by top ten test cases from technique A, suggesting that the techniques may be \textit{complimentary}. To evaluate the similarity between different TCP techniques, we utilize and build upon similarity analysis used in recent work \cite{Henard:ICSE16,Henard:TSE14} and construct binary vector representations of detected faults for each technique and then calculate the distance between these vectors as a similarity measure. 

We employ two methodologies in order to give a comprehensive view of the similarity of the studied TCPs.  At the core of both of these techniques is a measure of similarity using the Jaccard distance to determine the distance between vectorized binary representations of detected faults (where a 1 signifies a found fault and a 0 signifies an undiscovered fault) for different techniques across individual or groups of subject programs. We use the following definition \cite{Henard:ICSE16}:
\vspace{-0.2cm}
\begin{equation}\label{Equ:Jaccard}
    J(T_A^i,T_B^i)=\frac{\mid T_A^i \cap T_B^i \mid}{ \mid T_A^i \cup T_B^i \mid}
    \vspace{-0.2cm}
\end{equation}
where $T_A^i$ represents the binary vectorized discovered faults of some studied technique A after the execution of the $i^{th}$ test case in the techniques prioritized set, and $T_B^i$ represents the same meaning for some studied technique B and $0 \leq J(T_A^i, T_B^i) \leq 1$.  While we use the same similarity metric as in \cite{Henard:ICSE16}, we report two types of results: 1) results comparing the similarity of the studied static and dynamic techniques using the average Jaccard coefficient across all subjects at different test-case granularities, and 2) results comparing each technique in a pair-wise manner for each subject program. For the second type of analysis, we examine each possible pair of techniques and rank each subject program according to Jaccard coefficient as highly similar (1.0 - 0.75), similar (0.749 - 0.5), dissimilar (0.49 - 0.25), or highly dissimilar (0.249-0). This gives a more informative view of how similar two techniques might be for different subject programs. To construct both types of binary fault vectors, we use the same fault selection methodology used to calculate the APFD, that is, we randomly sample 500 faults from the set of known discoverable faults for each subject.

\textbf{RQ$_4$:} The final \textit{goal} of our study is to understand the efficiency of static techniques, in terms of execution costs. Note that, we only focus on the efficiency of static techniques, since dynamic techniques are typically run on the previous version of a program to collect coverage information, and thus the temporal overhead is quite high and well-studied. To evaluate the efficiency of static techniques, we collect two types of time information: the time for pre-processing and the time for prioritization. The time for pre-processing contains different phases for different techniques. For example, TP$_{cg-tot}$ and TP$_{cg-add}$ need to build the call graphs for each test case. TP$_{str}$ needs to analyze the source code to extract identifiers and comments for each test case. Besides, TP$_{topic}$ needs to pre-process extracted textual information and use the R-LDA package and Mallet \cite{Mallet} to build topic models. The time for prioritization refers to the time cost for TCP on different subjects.
\vspace{-0.1cm}
\subsection{Tools and Experimental Hardware}
\vspace{-0.05cm}
\label{sec:impl}
We reimplemented all of the studied dynamic and static TCPs in Java according to the specifications and descriptions in their corresponding papers, since the implementations were not available from the original authors and had to be adapted to our subjects. All three of the authors carefully reviewed and tested the code to make sure the reimplementation is reliable. We also invited an expert working in the area of test case prioritization to review our code.

\noindent
\textbf{TP$_{cg-tot}$/TP$_{cg-add}$:} Following the paper by Zhang \emph{et al.} \cite{Zhang:ICSM09}, we use the {\em IBM T. J. Watson Libraries for Analysis} (WALA) \cite{WALA} to collect the RTA static call graph for each test, and traverse the call graphs to obtain a set of relevant methods for each test case. Then, we implement two greedy strategies (i.e., total and additional) to prioritize test cases.

\noindent
\textbf{TP$_{str}$:} Based on the paper by Ledru \emph{et al.} \cite{Ledru:ASE12}, each test case is treated as one string without any preprocessing. Thus, we directly use JDT \cite{JDT} to collect the textual test information for each JUnit test, and then calculate the Manhattan distances between test cases to select the one that is farthest from the prioritized test cases.

\noindent
\textbf{TP$_{topic-r}$ and TP$_{topic-m}$:} Following the topic-based TCP paper \cite{Thomas:EMSE14}, we first use JDT to extract identifiers and comments from each JUnit test, and then pre-process those (e.g., splitting, removing stop words, and stemming). To build topic models, we used the R-LDA package \cite{LDA} for TP$_{topic-r}$ and Mallet \cite{Mallet} for TP$_{topic-m}$. All parameters are set with previously used values \cite{Thomas:EMSE14, Chen:13}. Finally, we calculated the Manhattan distances between test cases, and selected the ones that are farthest from the prioritized test cases.

\noindent
\textbf{Dynamic TCP techniques:} We use the ASM bytecode manipulation and analysis toolset~\cite{ASM} to collect the coverage information for each test. Specifically, in our empirical study, it obtains a set of statements that can be executed by each test method or test class. The greedy techniques are replicated based on the paper by Rothermel \emph{et al.} \cite{Rothermel:99}. For the ART and search-based techniques, we follow the methodology described in their respective papers \cite{Jiang:09,Li:07}.

\noindent
\textbf{Experimental Hardware:} The experiments were carried out on Thinkpad X1 laptop with Intel Core i5-4200 2.30 GHz processor and 8 GB DDR3 RAM.

\begin{figure*}[t]
\vspace{-0.2cm}
\center
\subfigure[The values of APFD on test-class level across all subject programs.\label{fig:apfd-tc}]{
\includegraphics[width=0.97\textwidth]{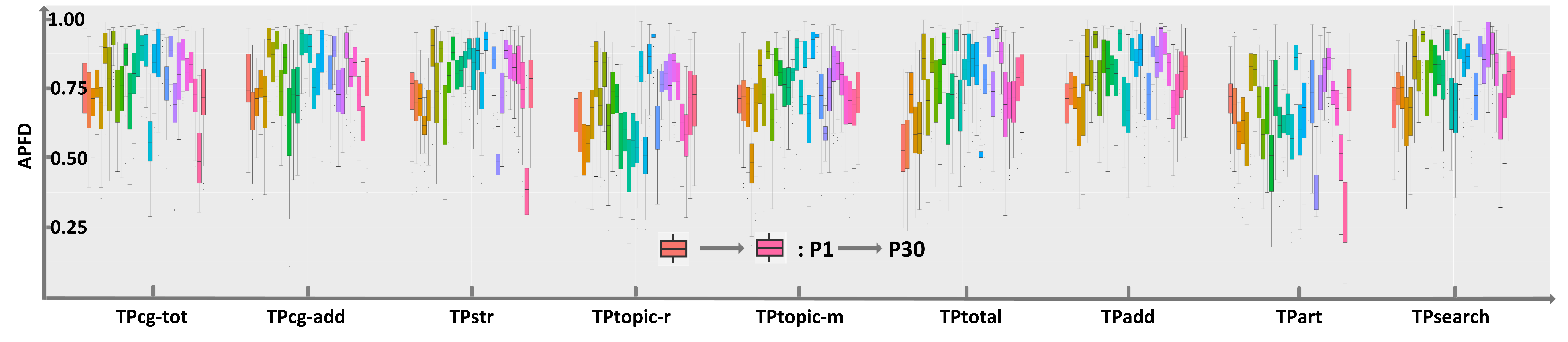}\vspace{-0.22cm}}\vspace{-0.25cm}
\subfigure[The values of APFD on test-method level across all subject programs.\label{fig:apfd-tm}]{
\includegraphics[width=0.97\textwidth]{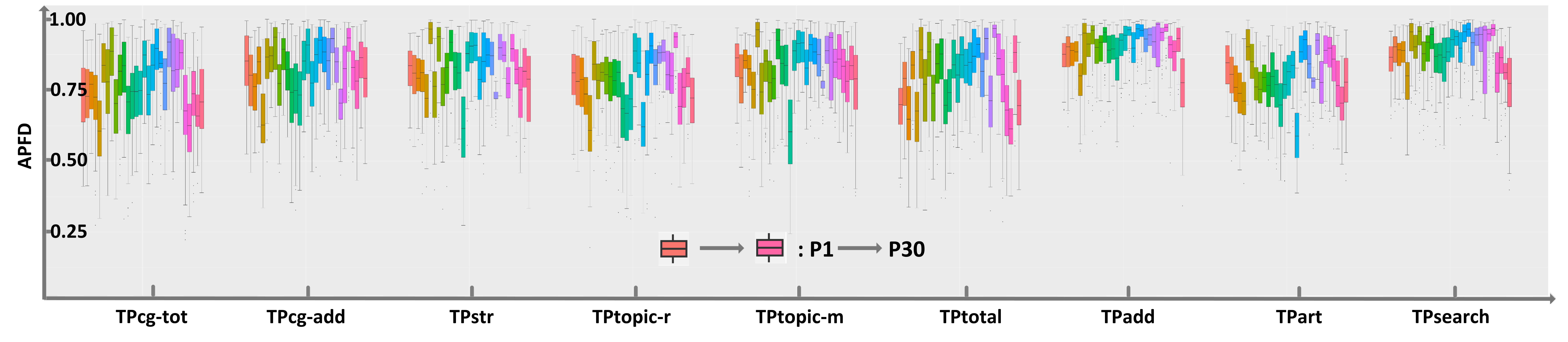}\vspace{-0.22cm}}
\vspace{-0.55cm}
\small
\caption{The box-and-whisker plots represent the values of APFD for different TCP techniques at different test granularities. The x-axis represents the APFD values. The y-axis represents the different techniques. The central box of each plot represents the values from the lower to upper quartile (i.e., 25 to 75 percentile). }
\vspace{-0.05cm}
\end{figure*}
\begin{table*}[t]
\center
\vspace{-0.5cm}
\scriptsize
\caption{\small Results for the ANOVA and Tukey HSD tests on the average APFD values depicted in Figures \ref{fig:apfd-tc} \& \ref{fig:apfd-tm}.\label{tab:stat}}
\begin{tabular}{|c|c|c|c|c|c|c|c|c|c|c|c|c|}
\hline
Granularity&Metric&{TP$_{cg-tot}$}&{TP$_{cg-add}$}&{TP$_{str}$}&{TP$_{topic-r}$}&{TP$_{topic-m}$}&{TP$_{total}$}&{TP$_{add}$}&{TP$_{art}$}&{TP$_{search}$}&p-value\\
\hline\hline
\multirow{2}{*}{Test-class}&Avg&0.782&0.793&0.769&0.680&0.747&0.748&0.789&0.659&0.786&\multirow{2}{*}{1.86e-7}\\\cline{2-11}
&HSD&A&A&A&BC&AB&AB&A&C&A&\\\hline\hline
\multirow{2}{*}{Test-method}&Avg&0.768&0.816&0.819&0.783&0.828&0.794&0.898&0.795&0.883&\multirow{2}{*}{1.69e-13}\\\cline{2-11}
&HSD&D&CD&CD&CD&BC&CD&A&CD&AB&\\\hline
\end{tabular}
\vspace{-0.6cm}
\end{table*}

\vspace{-0.25cm}
\section{Results}
\vspace{-0.05cm}
\label{sec:res}
In this section, we outline the experimental results to answer the \textbf{RQs} listed in Section~\ref{sec:study}.
\vspace{-0.05cm}
\subsection{RQ$_1$ \& RQ$_2$: Effectiveness of Studied Techniques at Different Granularities}
\vspace{0.35cm}
\begin{table*}[t]
\vspace{-0.2cm}
\center
\setlength{\tabcolsep}{2pt}
\tiny
\caption{\small The table shows the results of Wilcoxon signed rank test on the average APFD values for each pair of TCP techniques. The techniques T1 to T9 refer to TP$_{cg-tot}$, TP$_{cg-add}$, TP$_{str}$, TP$_{topic-r}$, TP$_{topic-m}$, TP$_{total}$, TP$_{add}$, TP$_{art}$, TP$_{search}$ respectively. For each pair of TCP techniques, there are two sub-cells. The first one refers to the p-value at test-class level and the second one refers to the p-value at test-method level. If a p-value is less than 0.05, the corresponding cell is shaded. \label{tab:stat-w}}
\begin{tabular}{|c||c|c||c|c||c|c||c|c||c|c||c|c||c|c||c|c||c|c|}
\hline
&\multicolumn{2}{c||}{T1}&\multicolumn{2}{c||}{T2}&\multicolumn{2}{c||}{T3}&\multicolumn{2}{c||}{T4}&\multicolumn{2}{c||}{T5}&\multicolumn{2}{c||}{T6}&\multicolumn{2}{c||}{T7}&\multicolumn{2}{c||}{T8}&\multicolumn{2}{c|}{T9}\\
\hline\hline
T1&\cellcolor[gray]{0.8}-&\cellcolor[gray]{0.8}-&0.06&\cellcolor[gray]{0.8}3.1e-04&0.61&\cellcolor[gray]{0.8}3.4e-03&\cellcolor[gray]{0.8}4.5e-04&0.32&\cellcolor[gray]{0.8}0.03&\cellcolor[gray]{0.8}1.1e-03&0.10&\cellcolor[gray]{0.8}0.02&0.59&\cellcolor[gray]{0.8}1.7e-06&\cellcolor[gray]{0.8}2.8e-04&0.06&0.70&\cellcolor[gray]{0.8}1.7e-06\\\hline
T2&0.06&\cellcolor[gray]{0.8}3.1e-04&\cellcolor[gray]{0.8}-&\cellcolor[gray]{0.8}-&0.24&0.89&\cellcolor[gray]{0.8}1.3e-04&\cellcolor[gray]{0.8}0.01&\cellcolor[gray]{0.8}0.02&0.39&\cellcolor[gray]{0.8}0.03&0.07&0.67&\cellcolor[gray]{0.8}2.6e-06&\cellcolor[gray]{0.8}2.4e-05&0.32&0.85&\cellcolor[gray]{0.8}1.5e-05\\\hline
T3&0.61&\cellcolor[gray]{0.8}3.4e-03&0.24&0.89&\cellcolor[gray]{0.8}-&\cellcolor[gray]{0.8}-&\cellcolor[gray]{0.8}2.3e-03&\cellcolor[gray]{0.8}5.0e-03&\cellcolor[gray]{0.8}0.01&\cellcolor[gray]{0.8}0.04&0.13&0.06&0.45&\cellcolor[gray]{0.8}2.9e-06&\cellcolor[gray]{0.8}1.1e-05&0.09&0.48&\cellcolor[gray]{0.8}7.7e-06\\\hline
T4&\cellcolor[gray]{0.8}4.5e-04&0.32&\cellcolor[gray]{0.8}1.3e-04&\cellcolor[gray]{0.8}0.01&\cellcolor[gray]{0.8}2.3e-03&\cellcolor[gray]{0.8}5.0e-03&\cellcolor[gray]{0.8}-&\cellcolor[gray]{0.8}-&\cellcolor[gray]{0.8}7.2e-04&\cellcolor[gray]{0.8}2.0e-03&\cellcolor[gray]{0.8}6.4e-03&0.47&\cellcolor[gray]{0.8}2.6e-05&\cellcolor[gray]{0.8}1.7e-06&0.81&0.22&\cellcolor[gray]{0.8}2.6e-05&\cellcolor[gray]{0.8}1.7e-06\\\hline
T5&\cellcolor[gray]{0.8}0.03&\cellcolor[gray]{0.8}1.1e-03&\cellcolor[gray]{0.8}0.02&0.39&\cellcolor[gray]{0.8}0.01&\cellcolor[gray]{0.8}0.04&\cellcolor[gray]{0.8}7.2e-04&\cellcolor[gray]{0.8}2.0e-03&\cellcolor[gray]{0.8}-&\cellcolor[gray]{0.8}-&0.70&\cellcolor[gray]{0.8}0.03&\cellcolor[gray]{0.8}3.9e-03&\cellcolor[gray]{0.8}3.2e-06&\cellcolor[gray]{0.8}4.5e-04&\cellcolor[gray]{0.8}0.02&\cellcolor[gray]{0.8}8.2e-03&\cellcolor[gray]{0.8}2.2e-05\\\hline
T6&0.10&\cellcolor[gray]{0.8}0.02&\cellcolor[gray]{0.8}0.03&0.07&0.13&0.06&\cellcolor[gray]{0.8}6.4e-03&0.47&0.70&\cellcolor[gray]{0.8}0.03&\cellcolor[gray]{0.8}-&\cellcolor[gray]{0.8}-&\cellcolor[gray]{0.8}6.7e-03&\cellcolor[gray]{0.8}4.3e-06&\cellcolor[gray]{0.8}0.01&0.69&\cellcolor[gray]{0.8}7.6e-03&\cellcolor[gray]{0.8}2.0e-05\\\hline
T7&0.59&\cellcolor[gray]{0.8}1.7e-06&0.67&\cellcolor[gray]{0.8}2.6e-06&0.45&\cellcolor[gray]{0.8}2.9e-06&\cellcolor[gray]{0.8}2.6e-05&\cellcolor[gray]{0.8}1.7e-06&\cellcolor[gray]{0.8}3.9e-03&\cellcolor[gray]{0.8}3.2e-06&\cellcolor[gray]{0.8}6.7e-03&\cellcolor[gray]{0.8}4.3e-06&\cellcolor[gray]{0.8}-&\cellcolor[gray]{0.8}-&\cellcolor[gray]{0.8}3.7e-05&\cellcolor[gray]{0.8}1.9e-06&0.14&\cellcolor[gray]{0.8}4.2e-04\\\hline
T8&\cellcolor[gray]{0.8}2.8e-04&0.06&\cellcolor[gray]{0.8}2.4e-05&0.32&\cellcolor[gray]{0.8}1.1e-05&0.09&0.81&0.22&\cellcolor[gray]{0.8}4.5e-04&\cellcolor[gray]{0.8}0.02&\cellcolor[gray]{0.8}0.01&0.69&\cellcolor[gray]{0.8}3.7e-05&\cellcolor[gray]{0.8}1.9e-06&\cellcolor[gray]{0.8}-&\cellcolor[gray]{0.8}-&\cellcolor[gray]{0.8}4.4e-05&\cellcolor[gray]{0.8}3.5e-06\\\hline
T9&0.70&\cellcolor[gray]{0.8}1.7e-06&0.85&\cellcolor[gray]{0.8}1.5e-05&0.48&\cellcolor[gray]{0.8}7.7e-06&\cellcolor[gray]{0.8}2.6e-05&\cellcolor[gray]{0.8}1.7e-06&\cellcolor[gray]{0.8}8.2e-03&\cellcolor[gray]{0.8}2.2e-05&\cellcolor[gray]{0.8}7.6e-03&\cellcolor[gray]{0.8}2.0e-05&0.14&\cellcolor[gray]{0.8}4.2e-04&\cellcolor[gray]{0.8}4.4e-05&\cellcolor[gray]{0.8}3.5e-06&\cellcolor[gray]{0.8}-&\cellcolor[gray]{0.8}-\\\hline
\end{tabular}
\end{table*}
\vspace{-0.45cm}

\begin{table*}[t]
\center
\tiny
\vspace{-0.65cm}
\setlength{\tabcolsep}{4.8pt}
\caption{\small The tables show the classification of subjects on different granularities using Jaccard distance. The four values in each cell are the numbers of subject projects, the faults of which detected by two techniques are highly dissimilar, dissimilar, similar and highly similar respectively. The technique enumeration is consistent with Table 4. \vspace{-0.3cm}}
\subtable[This table shows the classification of subjects at the cut point 10\% on test-class level. \vspace{-0.4cm}\label{tab:tcj-10}]{
\begin{tabular}{|c|p{1.2mm}p{1.2mm}p{1.2mm}p{1.2mm}|p{1.2mm}p{1.2mm}p{1.2mm}p{1.2mm}|p{1.2mm}p{1.2mm}p{1.2mm}p{1.2mm}|p{1.2mm}p{1.2mm}p{1.2mm}p{1.2mm}|p{1.2mm}p{1.2mm}p{1.2mm}p{1.2mm}|p{1.2mm}p{1.2mm}p{1.2mm}p{1.2mm}|p{1.2mm}p{1.2mm}p{1.2mm}p{1.2mm}|p{1.2mm}p{1.2mm}p{1.2mm}p{1.2mm}|p{1.2mm}p{1.2mm}p{1.2mm}p{1.2mm}|}
\hline
&\multicolumn{4}{c|}{T1}&\multicolumn{4}{c|}{T2}&\multicolumn{4}{c|}{T3}&\multicolumn{4}{c|}{T4}&\multicolumn{4}{c|}{T5}&\multicolumn{4}{c|}{T6}&\multicolumn{4}{c|}{T7}&\multicolumn{4}{c|}{T8}&\multicolumn{4}{c|}{T9}\\\hline
T1&--&--&--&--&1&2&11&16&6&7&7&10&18&8&1&3&9&6&8&7&7&9&4&10&7&7&7&9&19&7&3&1&7&8&6&9\\\hline
T2&1&2&11&16&--&--&--&--&6&7&7&10&17&10&2&1&8&12&5&5&6&10&8&6&7&8&7&8&18&10&2&0&7&8&6&9\\\hline
T3&6&7&7&10&6&7&7&10&--&--&--&--&16&6&6&2&5&7&7&11&10&6&8&6&8&7&9&6&17&7&2&4&8&7&8&7\\\hline
T4&18&8&1&3&17&10&2&1&16&6&6&2&--&--&--&--&9&7&7&7&16&7&4&3&17&6&5&2&17&6&4&3&17&7&3&3\\\hline
T5&9&6&8&7&8&12&5&5&5&7&7&11&9&7&7&7&--&--&--&--&10&11&3&6&10&10&4&6&16&8&4&2&10&10&4&6\\\hline
T6&7&9&4&10&6&10&8&6&10&6&8&6&16&7&4&3&10&11&3&6&--&--&--&--&0&5&7&18&16&8&3&3&0&5&8&17\\\hline
T7&7&7&7&9&7&8&7&8&8&7&9&6&17&6&5&2&10&10&4&6&0&5&7&18&--&--&--&--&13&9&5&3&0&0&1&29\\\hline
T8&19&7&3&1&18&10&2&0&17&7&2&4&17&6&4&3&16&8&4&2&16&8&3&3&13&9&5&3&--&--&--&--&13&8&6&3\\\hline
T9&7&8&6&9&7&8&6&9&8&7&8&7&17&7&3&3&10&10&4&6&0&5&8&17&0&0&1&29&13&8&6&3&--&--&--&--\\\hline
\hline
Total&74&54&47&65&70&67&48&55&76&54&54&56&127&57&32&24&77&71&42&50&65&61&45&69&62&52&45&81&129&63&29&19&62&53&42&83\\\hline
\end{tabular}
}
\subtable[This table shows the classification of subjects at the cut point 10\% on test-method level. \vspace{-0.45cm}\label{tab:tmj-10}]{
\begin{tabular}{|c|p{1.2mm}p{1.2mm}p{1.2mm}p{1.2mm}|p{1.2mm}p{1.2mm}p{1.2mm}p{1.2mm}|p{1.2mm}p{1.2mm}p{1.2mm}p{1.2mm}|p{1.2mm}p{1.2mm}p{1.2mm}p{1.2mm}|p{1.2mm}p{1.2mm}p{1.2mm}p{1.2mm}|p{1.2mm}p{1.2mm}p{1.2mm}p{1.2mm}|p{1.2mm}p{1.2mm}p{1.2mm}p{1.2mm}|p{1.2mm}p{1.2mm}p{1.2mm}p{1.2mm}|p{1.2mm}p{1.2mm}p{1.2mm}p{1.2mm}|}
\hline
&\multicolumn{4}{c|}{T1}&\multicolumn{4}{c|}{T2}&\multicolumn{4}{c|}{T3}&\multicolumn{4}{c|}{T4}&\multicolumn{4}{c|}{T5}&\multicolumn{4}{c|}{T6}&\multicolumn{4}{c|}{T7}&\multicolumn{4}{c|}{T8}&\multicolumn{4}{c|}{T9}\\\hline
T1&--&--&--&--&2&8&16&4&6&12&7&5&5&15&8&2&7&11&7&5&3&8&9&10&3&9&12&6&10&13&6&1&1&12&11&6\\\hline
T2&2&8&16&4&--&--&--&--&4&11&10&5&2&12&12&4&3&12&13&2&1&12&11&6&1&9&11&9&7&12&9&2&2&9&11&8\\\hline
T3&6&12&7&5&4&11&10&5&--&--&--&--&3&13&13&1&0&2&10&18&5&8&12&5&2&9&10&9&9&12&9&0&3&8&12&7\\\hline
T4&5&15&8&2&2&12&12&4&3&13&13&1&--&--&--&--&3&14&12&1&4&15&8&3&3&14&10&3&4&17&7&2&2&14&10&4\\\hline
T5&7&11&7&5&3&12&13&2&0&2&10&18&3&14&12&1&--&--&--&--&4&11&11&4&3&3&19&5&5&14&11&0&4&4&18&4\\\hline
T6&3&8&9&10&1&12&11&6&5&8&12&5&4&15&8&3&4&11&11&4&--&--&--&--&1&9&11&9&11&14&4&1&2&9&10&9\\\hline
T7&3&9&12&6&1&9&11&9&2&9&10&9&3&14&10&3&3&3&19&5&1&9&11&9&--&--&--&--&5&12&11&2&1&2&5&22\\\hline
T8&10&13&6&1&7&12&9&2&9&12&9&0&4&17&7&2&5&14&11&0&11&14&4&1&5&12&11&2&--&--&--&--&6&12&10&2\\\hline
T9&1&12&11&6&2&9&11&8&3&8&12&7&2&14&10&4&4&4&18&4&2&9&10&9&1&2&5&22&6&12&10&2&--&--&--&--\\\hline\hline
Total&37&88&76&39&22&85&93&40&32&75&83&50&26&114&80&20&29&71&101&39&31&86&76&47&19&67&89&65&57&106&67&10&21&70&87&62\\\hline
\end{tabular}
}
\vspace{-0.8cm}
\end{table*} 

 The values of APFD across all subjects at class-level are shown in Figure~\ref{fig:apfd-tc} and Table \ref{tab:stat}.  Based on this figure, we make the following observations.  First, somewhat surprisingly, at the test-class level, the static TP$_{cg-add}$ technique performs the best across all studied TCP techniques (including all dynamic techniques) with an average APFD value of 0.793 (see Table~\ref{tab:stat}). TP$_{cg-tot}$ performs worse than TP$_{cg-add}$, followed by TP$_{str}$, TP$_{topic-m}$ and TP$_{topic-r}$.  The best performing dynamic technique at class-level is TP$_{add}$ followed by  TP$_{search}$, TP$_{total}$, and TP$_{art}$.  It is notable that at test-class level granularity, the most effective dynamic technique TP$_{add}$ and the most effective static technique perform similarly, 0.793 versus 0.789 respectively.  This suggests that at the test-class level, the call-graph based strategy performs about as well as dynamic coverage information, which is notable. Additionally, overall the static techniques outperform the dynamic techniques at test-class granularity.

To further investigate \textbf{RQ$_1$} and answer \textbf{RQ$_2$} we ran all of the subject TCPs on the subject programs at test-method level which we can compare to the results at test-class level outlined above (see \text{RQ$_1$}). The results are shown in Figure~\ref{fig:apfd-tm} and Table~\ref{tab:stat}. First, when examining the static techniques \textit{with test-method granularity}, they perform differently as compared to the results on test-class level. Surprisingly, T$_{topic-m}$ (0.828) performs better than the other static techniques, followed by TP$_{str}$, TP$_{cg-add}$, TP$_{topic-r}$ and TP$_{cg-tot}$ respectively. It is worth noting that the effectiveness of the topic-model based technique varies quite dramatically depending on the tools used for its implementation: Mallet \cite{Mallet} significantly outperforms the R-based implementation.  However, as a whole the effectiveness of the dynamic techniques outpaces that of the static techniques at method-level granularity, with TP$_{add}$ performing the best of all studied techniques (0.898).  This finding is consistent with previous studies \cite{Hao:TOSEM14}. Overall, on average, all static and dynamic TCPs perform better on test-method level as compared to the results on test-class level. Logically, this is not surprising, as using a finer level of granularity (e.g., prioritizing individual test-methods) gives each technique more flexibility, which leads to more accurate targeting and prioritization. Furthermore, the ranges of average of APFD for all TCPs on test-method level are smaller than the results on test-class level, confirming that the performance of the TCP at test-method level is more stable.

The ranges of APFD values reflect the robustness of the TCPs. For example, the range of average of APFD across all subjects at test-class level for TP$_{add}$ is the smallest (i.e.,0.612-0.919), implying that the performance of TP$_{add}$ is usually stable despite differing types of subjects. Conversely, the ranges of APFD values for TP$_{str}$ and TP$_{art}$ are much larger (0.391-0.917 for TP$_{str}$, 0.298-0.852 for TP$_{art}$), implying that their performance varies on different types of subjects.

To further investigate the finding that static techniques tend to have a higher variance in terms of effectiveness depending on the program type, we investigated further by inspecting several subject programs. One illustrative example is that {\em scribe-java} scores $0.646$ and $0.606$ for the average values of APFD under TP$_{str}$ and TP$_{topic-r}$ respectively, which are notably worse than the results of TP$_{cg-tot}$ (0.718) and TP$_{cg-add}$ (0.733). To understand the reason for this discrepancy, we analyzed the test code and found that {\em Scribe-java} is documented/written more poorly than other programs. For instance, the program uses meaningless comments and variable names such as `$param1$', `$param2$', `$v1$', `$v2$' etc.  This confirms the previously held notion \cite{Thomas:EMSE14} that static techniques which aim to prioritize test-cases through text-based diversity metrics experience performance degradation when applied to test cases written in a poor/generic fashion.

Finally, to check for statistically significant variations in the mean APFD values across all subjects and confirm/deny our null hypothesis for \textbf{RQ$_1$}, we performed a one-way ANOVA and Tukey HSD test.  The results of the ANOVA test, given in the last column of Table \ref{tab:stat}, are well below our established threshold of 0.05, thus signifying that the subject programs are statistically different from one another.  This rejects the null hypothesis and we conclude that there are statistically significant differences between different TCP techniques at the differing granularities.  The results of the Tukey HSD test illustrate the statistically significant differences between the static and dynamic techniques, by grouping the techniques into categories with \textit{A} representing the best performance and \textit{D} representing the worst.  For test-class level, we see that the groupings slightly favor the static techniques, as more of them are grouped in the top-ranked \textit{A} category.  For test method level, it is clear that the dynamic techniques outperform the static, as far more dynamic techniques are ranked in the better performing categories.  In order to illustrate the individual relationships between strategies, we present the results of the Wilcoxon signed rank test for all pairs of techniques in Table \ref{tab:stat-w}.  The shaded cells represent the results that indicate a statistically significant difference between techniques across all the subjects (e.g., $p < 0.05$) In summary we answer \textbf{RQ$_1$} \& \textbf{RQ$_2$} as follows: 

\vspace{-0.15cm}
\begin{framed}{\vspace{-0.1cm}
{\bf RQ$_1$:} There is a statistically significant difference between the effectiveness of the studied techniques. On average, static technique TP$_{cg-add}$ is the most effective technique at test-class level, whereas dynamic technique TP$_{add}$ is the most effective technique at test-method level. Overall, the static techniques outperform the dynamic ones at test-class level, but the dynamic techniques outperform the static ones at test-method level.\vspace{-0.1cm}}
\end{framed}
\vspace{-0.25cm}
\begin{framed}{\vspace{-0.1cm}
{\bf RQ$_2$: } The test granularity significantly impacts the effectiveness of TCP techniques. All the studied techniques perform better at test-method level compared to {\bf RQ$_2$:$_{(continued)}$} test-class level. There is also less variation in the APFD values at method-level compared to class-level, which signifies that the performance of the studied techniques is more stable at test-method level.\vspace{-0.1cm}}
\end{framed}
\vspace{-0.2cm}

\vspace{-0.3cm}
\subsection{Similarity between Uncovered Faults for Different TCP techniques}
\begin{figure}[tb]
\centering
\vspace{-0.2cm}
\subfigure[\vspace{-0.2cm}Class-level results]{\hspace{-0.2cm}
\includegraphics[width=1.01\columnwidth]{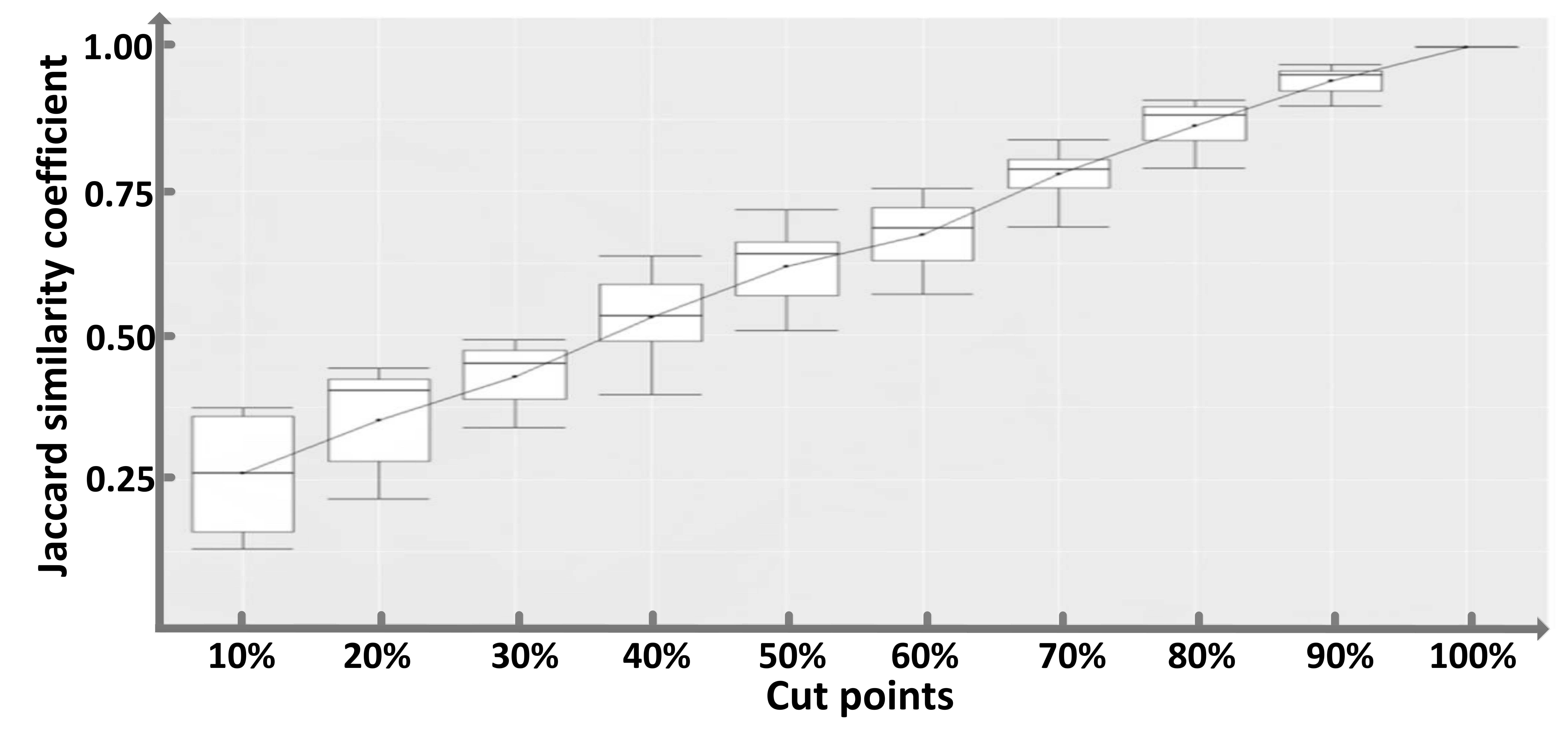}\label{fig:jac-c}}\vspace{-0.32cm}
\subfigure[Method-level Results]{\hspace{-0.2cm}
\includegraphics[width=1.01\columnwidth]{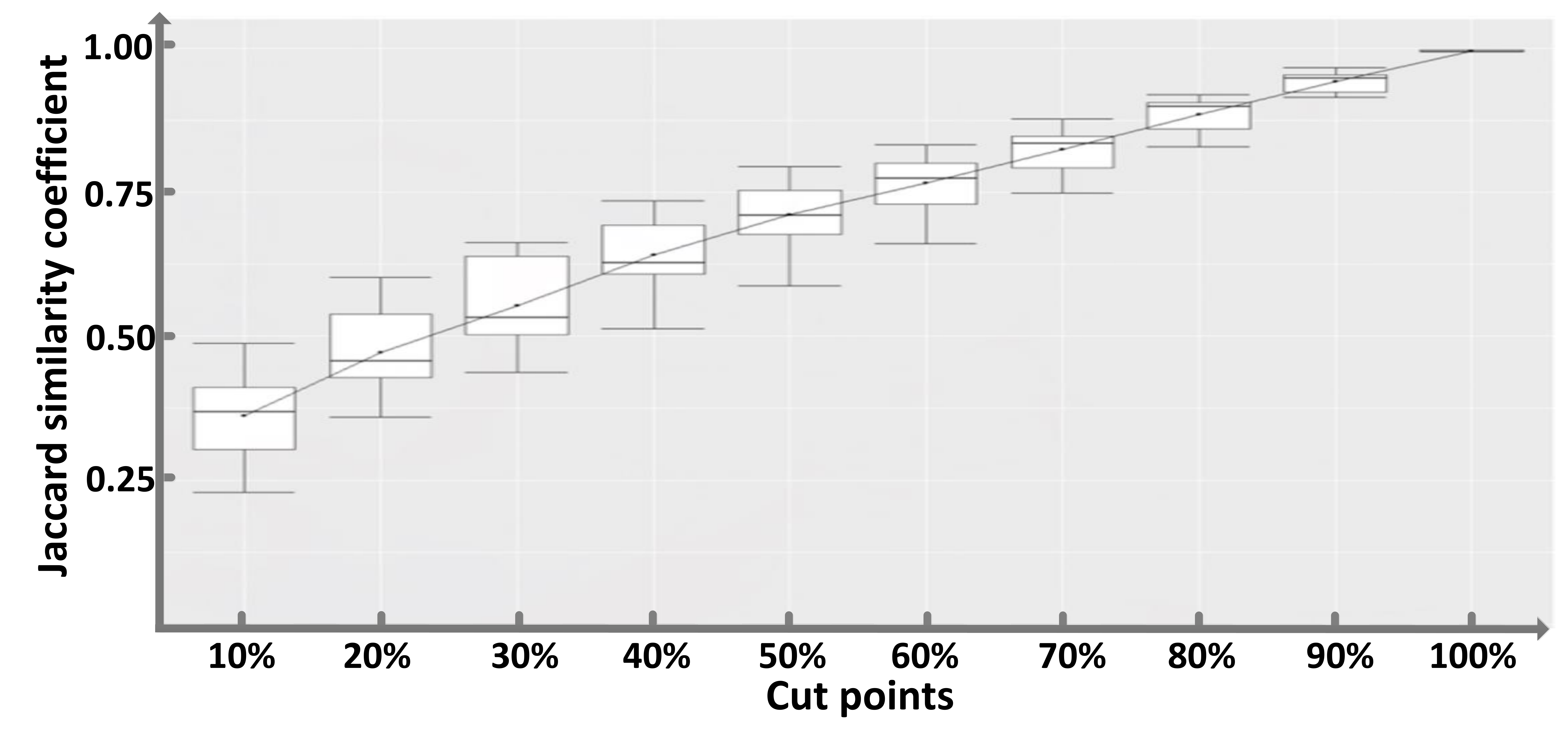}\label{fig:jac-m}}
\vspace{-0.7cm}
\caption{Average Jaccard similarity of faults detected between static and dynamic techniques across all subjects at method and class-level granularity.}
\vspace{-0.63cm}
\end{figure}

\vspace{-0.1cm}
The results for the similarity are shown in Tables \ref{tab:tcj-10} \& \ref{tab:tmj-10} and Figures \ref{fig:jac-m} \& \ref{fig:jac-c}. The two figures represent the results comparing the average Jaccard similarity of the studied static techniques to the studied dynamic techniques for all subject programs across 500 randomly sampled faults at different prioritization cut points. \Comment{Thus, this represents a high-level comparison of the similarity of the studied static techniques, against the studied dynamic techniques.}These results indicate that there is only a small amount of similarity between these two classifications of techniques at the higher level cut points. More specifically, for test-method level, only $\approx$ 30\% of the detected faults are similar between the two types of techniques for the top 10\% of the prioritized test cases, and at test-class level only about $\approx$ 25\% are similar for the top 10\% of prioritized test cases. This result illustrates one of the key findings of this study: The studied static and dynamic TCP techniques do not uncover similar program faults at the top cut points of prioritized test cases. The potential reason for these results is that different techniques use different types of information to prioritize test cases. For example, the studied static techniques typically aim to promote diversity between prioritized test cases using similarity/diversity metrics such as textual distance or call-graph information.  In contrast, the studied dynamic TCPs consider statement-level dynamic coverage to prioritize test cases. This finding raises interesting questions for future work regarding the possibility of combining static and dynamic information and the relative \textit{importance} of faults that differing techniques might uncover. It should be noted that different coverage granularities for dynamic TCPs may also effect the results of similarity, however we leave such an investigation for future work. From these figures we can also conclude that the techniques are slightly more similar at method level than at class level.

To further illustrate this point we calculated the Jaccard coefficients for each pair of TCPs for each subject program, and show the results in Table \ref{tab:tcj-10}. For each pair of techniques we group the subjects into the categories described in Section \ref{sec:study}.  Due to space limitations, we only show results for the top 10\% of prioritized test-cases, a complete dataset can be found at \cite{Qi:FSE16}. The results confirm the conclusions drawn from Figures \ref{fig:jac-c} \& \ref{fig:jac-m}. It is clear that when comparing the studied static and dynamic techniques, more subjects are classified into the \textit{highly-dissimilar} and \textit{dissimilar} categories. Another relevant conclusion that can be made is that the dissimilarity between techniques is not universal across all subjects. That is, even though two techniques may be dissimilar across several subjects, there are some cases where similarity still exists.  This suggests that only certain types of programs that exhibit different characteristics may present the opportunity of performance improvement for TCPs by using both static and dynamic information.

\begin{table*}[t]
\centering
\scriptsize
\vspace{-0.25cm}
\hspace{-1cm}
\caption{Execution costs for the static TCP techniques. The table lists the average, min, max, and sum of costs across all subject programs for both test-class level and test-method level (i.e., cost at test-class level/cost at test-method level). Time is measured in second. \label{tab:time-sum}}
\vspace{-0.05cm}
\begin{tabular}{|c|cccc|cccc|}
\hline
\multirow{2}{*}{Techniques}&\multicolumn{4}{c|}{Pre-processing}&\multicolumn{4}{c|}{Test Prioritization}\\
&Avg.&Min&Max&Sum&Avg.&Min&Max&Sum\\
\hline
TP$_{cg-tot}$&4.66/4.66&1.69/1.69&11.95/11.95&139.69/139.69&0.21/0.50&0/0&3.10/10.58&6.22/15.00\\
TP$_{cg-add}$&4.66/4.66&1.69/1.69&11.95/11.95&139.69/139.69&0.19/0.82&0/0&2.87/19.98&5.70/24.57\\
TP$_{str}$&0.40/0.41&0.06/0.08&2.95/2.41&12.05/12.35&3.57/1,960.20&0.02/0.02&67.13/57,134.30&107.25/58,805.94\\
TP$_{topic-r}$&0.50/0.53&0.11/0.11&3.19/3.74&14.94/15.86&0.14/1,362.52&0/0.02&1.38/40,594.66&4.10/40,875.75\\
TP$_{topic-m}$&0.72/4.28&0.13/0.22&6.18/50.01&21.66/128.48&0.16/373.52&0/0.09&1.68/10,925.26&4.83/11,205.73\\\hline
\end{tabular}
\vspace{-0.6cm}
\end{table*}

\vspace{-0.25cm}
\begin{framed}{\vspace{-0.15cm}
{\bf RQ$_3$: } The studied static and dynamic TCP techniques tend to discover dissimilar faults for the most highly prioritized test cases. Specifically, at the test-method level static and dynamic techniques agree only on $\approx$ 30\% of uncovered faults for the top 10\% of prioritized test cases. Additionally, a subset of subjects exhibit higher levels of ucovered fault similarity, suggesting that only software systems with certain characteristics may benefit from differing TCP approaches.\vspace{-0.15cm}}
\end{framed}
\vspace{-0.35cm}

\subsection{Efficiency of Static TCP Techniques}
The results of time costs for the studied static techniques at both of test-method and test-class levels are shown in Table~\ref{tab:time-sum}. Note that, the time of pre-processing for TP$_{cg-tot}$ and TP$_{cg-add}$ are the same for both method and class levels. As the table shows, all studied techniques require similar time to pre-process the data at both method and class levels and to rank test cases on class level. But the times for prioritization are quite different at method level. We find that TP$_{cg-tot}$ and TP$_{cg-add}$ take much less time to prioritize test cases (totaling 15.00 seconds and 24.57 seconds), as compared to TP$_{str}$ (totalling 58,805.94 seconds), TP$_{topic-r}$ (totalling 40,875.75 seconds) and TP$_{topic-m}$ (totalling 11,205,73 seconds). Specially, these two techniques take much longer time on some subjects (e.g., {\em P27} and {\em P30} ). However, these subjects have a large number of test cases (see Table~\ref{tab:sub}), implying that TP$_{str}$, TP$_{topic-r}$ and TP$_{topic-m}$ will take more time as the number of test cases increases. Overall, all techniques take a reasonable amount of time to preprocess data and prioritize test cases. At test-method level, TP$_{cg-tot}$ and TP$_{cg-add}$ are much more efficient. TP$_{str}$, TP$_{topic-r}$ and TP$_{topic-m}$ require more time to prioritize increasing numbers of test cases, answering \textbf{RQ$_4$}.
\vspace{-0.45cm}
\begin{framed}{
{\bf RQ$_4$: } On test-method level, TP$_{cg-tot}$ and TP$_{cg-add}$ are much more efficient in prioritizing test cases. TP$_{str}$, TP$_{topic-r}$ and TP$_{topic-m}$ would take more time when the number of test cases increases. The time of pre-processing and prioritization on test class level for all static techniques are quite similar.}
\end{framed}
\vspace{-0.15cm}

\vspace{-0.3cm}
\section{Threats to Validity}
\label{sec:threats}

\noindent
\textbf{Threats to Internal Validity:}
In our implementation, we used PIT to generate mutation faults to simulate real program faults. One potential threat is that the mutation faults may not reflect all ``natural" characteristics of real faults. However, mutation faults have been widely used in the domain of software engineering research and has been demonstrated to be representative of the actual program faults~\cite{Just:FSE14}. Additionally, further threats related to mutation testing include the potential bias introduced by equivalent and trivial mutants.  To mitigate these threats, we randomly selected 500 faults for each subject system when conducting our study related to the effectiveness and similarity of faults uncovered for the various techniques.  This follows the guidelines and methodology of previous well-accepted studies \cite{zhang2013bridging,Lu:ICSE16}, minimizing this threat.

To perform the study we reimplemented eight TCP techniques presented in prior work. It is possible that there may be some slight differences between the original authors' implementations and our own. However, we performed this task closely following the technical details of the prior techniques and set parameters following the guidelines in the original works.  Additionally, the authors of this paper met for and open code review related to the implementation of the studied approaches, and our implementation was reviewed by an expert in the field of test-case prioritization. Furthermore, based on our general findings, we believe our implementations to be accurate.

\noindent
\textbf{Threats to External Validity:} The main external threat to our study is that we experimented on only 30 software systems, which may impact the generalizability of the results. Involving more subject programs would make it easier to reason about how the studied TCP techniques would perform on software systems of different languages and purposes. However, we chose 30 systems with varying sizes (3,2KLoC - 83 KLoC) and different numbers of detectable faults (132 - 20,957), which makes for a highly representative set of Java programs.  Additionally, some subjects were used as benchmarks in recent papers~\cite{Saha:ICSE15}. Thus, we believe our study parameters have sufficiently mitigated this threat to a point where useful and actionable conclusions can be drawn in the context of our research questions.

Finally, we selected four static TCP techniques to experiment with in our empirical study. There are some other recent works proposing static TCP techniques~\cite{Arafeen:ICST13,Saha:ICSE15}, but we focus only on those which do not require additional inputs, such as code changes or requirements in this empirical study. Furthermore, we only compared the static techniques with four state-of-art dynamic TCP techniques with statement-level coverage. We do not study the potential impact of different coverage granularities on dynamic TCPs. However, these four techniques are highly representative of dynamic techniques and have been widely used in TCP evaluation \cite{Lu:ICSE16, Rothermel:99, Do:04}, and statement-level coverage has been shown to be \textit{at least as effective} as other coverage types \cite{Lu:ICSE16}. 

\vspace{-0.3cm}
\section{Lessons Learned}
\vspace{-0.1cm}
In this section we comment on the lessons learned from this study and their potential impact on future research:

\noindent
\textbf{Lesson 1.} Our study illustrates that different test granularities impact the effectiveness of TCP techniques, and that the finer, method-level, granularity achieves better performance in terms of APFD, detecting regression faults more quickly.  This finding should encourage researchers and practitioners to use method-level granularity, and perhaps explore even smaller granularities for regression test-case prioritization. Additionally, researchers should evaluate their newly proposed approaches on different test granularities to better understand the effectiveness of new approaches.

\noindent
\textbf{Lesson 2.} The performance of different TCPs varies across different subject programs. One technique may perform better on some subjects but perform worse on other subjects. For example, TP$_{topic}$ performs better than TP$_{cg-add}$ on $webbit$, but performs worse than TP$_{cg-add}$ on $wsc$. This finding suggests that the characteristics of each subject are important to finding suitable TCPs. Furthermore, we find that the selection of subject programs and the selection of implementation tools may carry a large impact regarding the results of the evaluation for TCPs (e.g., there can be large variance in the performance of different techniques depending on the subject, particularly for static approaches). This finding illustrates that the researchers need to evaluate their newly proposed techniques on a \textit{large} set of real subject programs to make their evaluation reliable. To facilitate this we provide links to download our subject programs and data at \cite{Qi:FSE16}.Additionally, a potential avenue for future research may be an adaptive TCP technique that is able to analyze certain characteristics of a subject program (e.g., complexity, test suite size, libraries used) and modify the prioritization technique to achieve peak performance.

\noindent
\textbf{Lesson 3.} Our findings illustrate that the studied static and dynamic TCP techniques agree on only a small number of found faults for the top ranked test-methods and classes ranked by the techniques. This suggests several relevant avenues for future research. For instance, (i) it may be useful to investigate specific TCP techniques to detect important faults faster when considering the fault severity/importance \cite{Elbaum:01,Varun:ICJST10,Kavitha:10} during testing; (ii) differing TCP techniques could be used to target specific types of faults or even faults in specific locations of a program; and (iii) static and dynamic information could \textit{potentially} be combined in order to achieve higher levels of effectiveness. Furthermore, the similarity study performed in this paper has not been a core part of many TCP evaluations, and we assert that such an analysis should be encouraged moving forward. While APFD gives a clear picture of the relative effectiveness of techniques, it cannot effectively illustrate \textit{the difference set} of detected faults between two techniques.  This is a critical piece of information when attempting to understand new techniques and how they relate to existing research. 

\vspace{-0.25cm}
\section{Conclusion}
\vspace{-0.05cm}
\label{sec:conclude}

In this work, we perform an extensive study empirically comparing the effectiveness, efficiency, and similarity of detected faults for static and dynamic TCP techniques on 30 real Java programs. The experiments were conducted at both test-method and test-class levels to understand the impact of different test granularities on the effectiveness of TCP techniques. The results indicate that the studied static techniques tend to outperform the studied dynamic techniques at the test-class level, whereas dynamic techniques tend to outperform the static techniques at test-method level. Additionally, we found that the faults uncovered by static and dynamic techniques for the highest prioritized test cases uncover mostly dissimilar faults, which suggests several promising avenues for future work. Finally, we found evidence suggesting that different TCP techniques tend to perform differently on different subject programs, which suggests that certain program characteristics may be important when considering which type of TCP technique to use.

\vspace{-0.25cm}
\section{Acknowledgments}
\vspace{-0.05cm}
We would like thank the anonymous FSE reviewers for their insightful comments that significantly improved this paper. We also would like thank Lingming Zhang for aiding us with the collection of the coverage information and for reviewing the implementations of our TCP techniques. This work is supported in part by the NSF CCF-1218129 and NSF CNS-1510239 grants. Any opinions, findings, and conclusions expressed herein are the authors' and do not necessarily reflect those of the sponsors.

\balance
\bibliographystyle{abbrv}
\bibliography{ms}

\end{document}